\input{aipcheck}

\documentclass[
    ,final            
  ]
  {aipproc}

\layoutstyle{8x11single}

\begin{document}

\title{The nucleon-nucleon system in chiral effective theory}


\author{Daniel R. Phillips}{
  address={Department of Physics and Astronomy, Ohio University, Athens, OH 45701}
}

\begin{abstract}
I discuss the conditions under which the application of chiral perturbation theory to the NN potential gives reliable results for NN scattering phase shifts. $\chi$PT also yields a convergent expansion for the deuteron charge operator. For cutoffs $< 1$ GeV, this produces precise predictions for deuterium's quadrupole and charge form factors in the range $Q^2 < 0.25$ GeV$^2$. 
 \end{abstract}

\maketitle


\section{Introduction}

The use of chiral perturbation theory ($\chi$PT) to compute nuclear forces is a problem that has received much attention over the past twenty years. This effort began with the seminal papers of Weinberg~\cite{We90} and is referred to in this contribution as "chiral effective theory" ($\chi$ET). $\chi$ET uses $\chi$PT to compute the NN potential up to some fixed order, $n$, in the chiral expansion in powers of $P \equiv (p,m_\pi)/\Lambda_{0}$. (Here the breakdown scale $\Lambda_{0}$ is nominally $m_\rho \sim 4 \pi f_\pi$, but in reality is somewhat lower for reactions involving baryons.) This NN potential is then iterated,  using the Schr\"odinger or Lippmann-Schwinger equation, to obtain the scattering amplitude. The resulting wave functions can be combined with, e.g. electromagnetic current operators, also derived from $\chi$PT, in order to derive results for few-nucleon-system observables that are grounded in QCD's pattern of chiral-symmetry breaking. In addition, $\chi$ET results can be assessed for convergence as a function of the order, $n$, of the calculation. This enables specification of theoretical uncertainties based on the anticipated size of the $O(P^{n+1})$ corrections to the process of interest. 

\section{NN scattering in $\chi$ET}

To solve the Schr\"odinger equation with potentials derived from $\chi$PT we must introduce a cutoff, $\Lambda$, on the intermediate states, because $\chi$PT potentials grow with momenta. The low-energy constants (LECs) multiplying contact interactions in the nucleon-nucleon part of the chiral Lagrangian should then be adjusted to eliminate any $\Lambda$-dependence in the effective theory's predictions for low-energy observables. If this is not possible we conclude that $\chi$ET is unable to give reliable predictions. 

At leading order (LO) the $\chi$ET potential contains a zero-derivative contact interaction that is operative only in NN partial waves with $L=0$. One-pion exchange is also present in the LO potential, and is active in all partial waves. It couples $S=1$ partial waves with $\Delta L=2$ through its tensor part. Ref.~\cite{Be02} showed that the LO $\chi$ET potential leads to a $\Lambda$-independent amplitude in the limit $\Lambda \rightarrow \infty$ in the ${}^3$S$_1$-${}^3$D$_1$ channel. At fixed pion mass a $\Lambda$-independent result is also obtained in the ${}^1$S$_0$ channel~\cite{Be02}. Recently we showed how to "subtractively renormalize" the LO equations for NN scattering in these two channels~\cite{Ya08}. This technique eliminates the contact interaction in favor of a low-energy observable (e.g. the relevant NN scattering length). This makes it straightforward to take the limit $\Lambda \rightarrow \infty$: no fine-tuning of the pertinent LEC is necessary.  The resulting phase shifts (and one mixing parameter) do not provide anything like a precision description of NN data, but they are, at least, a well-defined, renormalized LO calculation. 

In Refs.~\cite{Or96,Ep99} (Refs.~\cite{EM03,Ep05}) $V$ was computed to $O(P^2)$ and $O(P^3)$ ($O(P^4)$), and the several NN LECs which appear in $V$ were fitted to NN data for a range of cutoffs between 500 and 800 MeV. The $O(P^4)$ predictions contain very little residual cutoff dependence in this range of $\Lambda$'s, and describe NN data with considerable accuracy. 

\begin{figure}[th]
\includegraphics[width=4.2in]{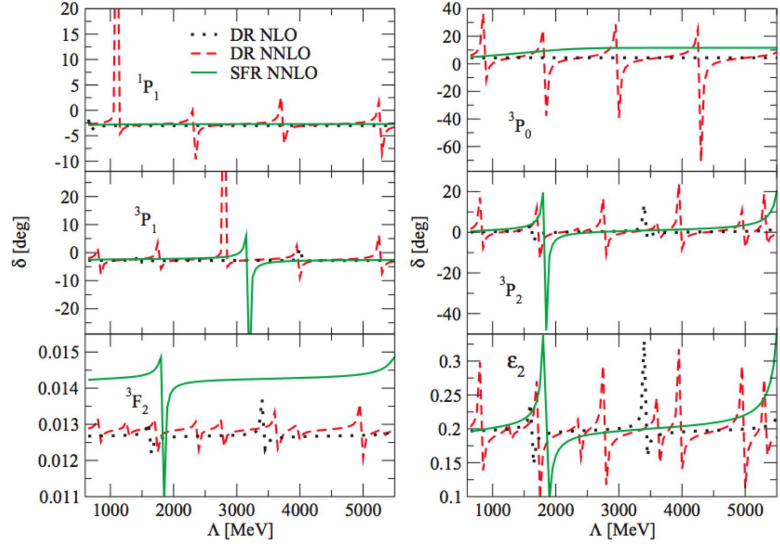}
\caption{P-wave NN phase shifts at $T_{\rm lab}=10$ MeV as a function of the Lippmann-Schwinger equation cutoff $\Lambda$. The black-dotted line is the $O(P^2)$ $\chi$ET calculation with the potential obtained via dimensional regularization. The dashed red (solid green) line is for the $O(P^3)$ potential calculated using dimensional regularization (spectral-function regularization). From Ref.~\cite{Ya09A}.}
\label{fig-Pphaseshifts}
\end{figure}

However, several recent papers showed that a LO $\chi$ET calculation does not yield stable predictions in partial waves with $L > 0$ once $\Lambda$ is sufficiently large~\cite{ES03,NTvK05,Bi06,PVRA06}. This problem occurs because only one-pion exchange is present in these waves at LO. The resulting singular potential has no NN LEC that permits renormalization. Recently we examined this problem within the context of subtractive renormalization. We confirmed the conclusion of Ref.~\cite{NTvK05}, i.e. any partial wave with $L > 0$ where one-pion exchange is attractive does not have stable LO results in $\chi$ET. We also showed that this problem is not removed at $O(P^2)$ or $O(P^3)$. In particular, at $O(P^3)$ two-pion exchange produces a highly singular, attractive potential. The  NN contact interactions needed to renormalize this potential are not present, e.g. in the ${}^3$P$_2$-${}^3$F$_2$ channel. The resulting lack of stability with $\Lambda$ of the NN phase shifts can be seen in Fig.~\ref{fig-Pphaseshifts}. We identified the cutoff where such difficulties first occur as $\approx 1$ GeV. 

\begin{figure}[thb]
\includegraphics[width=4in]{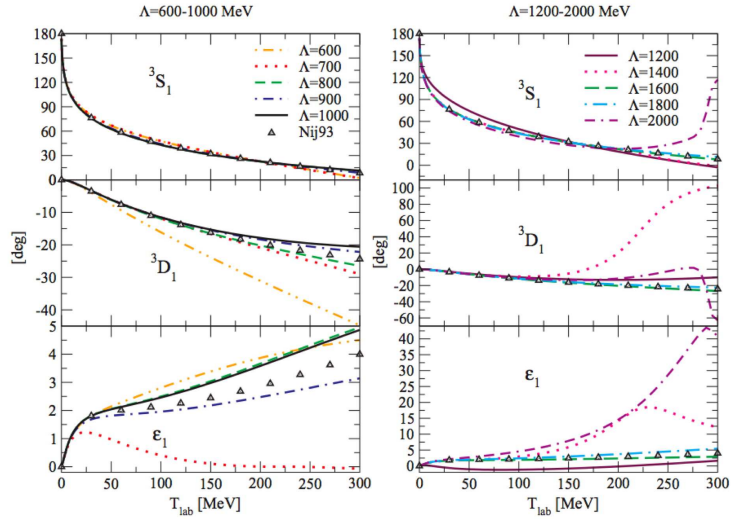}
\caption{The phase shifts associated with NN scattering in the $J=1$, $S=1$ channel at various cutoffs $\Lambda$. The triangles are the 1993 Nijmegen PSA~\cite{St93}. These results are for the $O(P^3)$ dimensionally regularized $\chi$ET potential. From Ref.~\cite{Ya09B}.}
\label{fig-tripphaseshifts}
\end{figure}

In Ref.~\cite{Ya09B} we used subtractive renormalization to calculate S-wave NN phase shifts from $O(P^2)$ and $O(P^3)$ $\chi$ET potentials. Here too we found that the phase shifts are not stable once $\Lambda > 1$ GeV (see Fig.~\ref{fig-tripphaseshifts}). In this case part of the problem is that the momentum-dependent contact interaction that appears at $O(P^2)$ has limited effect as $\Lambda \rightarrow \infty$~\cite{Wigner}. 

All of this indicates that $\chi$ET as formulated above is not properly renormalized, i.e. the impact of short-distance physics on the results is not under control. Thus the NN potential obtained by straightforward application of $\chi$PT cannot be used over a wide range of cutoffs. It can perhaps be used if we employ $\Lambda$'s in the vicinity of $m_\rho$~\cite{Le97,EM06}. Since the short-distance physics of the effective theory for $p \gg m_\rho$ is different to the short-distance physics of QCD itself, it is not clear that considering $\Lambda \gg m_\rho$ yields any extra information about the real impact of short-distance physics on observables. Using low cutoffs also has the advantage that relevant momenta are demonstrably within the domain of validity of $\chi$PT. This justifies $\chi$ET as a systematic theory, but at the cost of limiting us to $m_\pi \ll \Lambda < m_\rho$.

\section{Predictions for elastic electron-deuteron scattering at low $Q^2$}

Chiral effective theory can also be used to organize 
electromagnetic current operators as an expansion in powers of $P$. The pertinent matrix elements are then constructed via:
\begin{equation}
{\cal M}_\mu=\langle \psi^{(f)}| \sum_{k=0}^n J^{(k)}_\mu|\psi^{(i)} \rangle,
\end{equation}
where $|\psi^{(i)} \rangle$ and $|\psi^{(f)} \rangle$ are solutions of the Schr\"odinger equation for the $\chi$ET potential $V$---in principle at order $n$. 
(For recent progress towards the completely consistent construction of $V$ and $J_\mu$ see Refs.~\cite{EK09,Pa09}.)

Here we focus on deuteron charge and quadrupole form factors, calculated as matrix elements of the NN charge operator $J_0$ using
the standard (Breit-frame) formulae~\cite{Ph03}.  We compare $\chi$ET predictions for
$G_C$ and $G_Q$ with extractions from data for the deuteron
structure function $A$ and the tensor-polarization observable
$T_{20}$~\cite{data}. (For $\chi$ET calculations of the deuteron magnetic form factor, $G_M$, and deuteron photodisintegration see Refs.~\cite{Ph03,MW01,Pa99,Ro10}.)

The leading piece of $J_0^{(s)}$ occurs at $O(e)$:
$\langle {\bf p}'|J_0^{(s)}({\bf q})|{\bf p}\rangle=
|e|  \delta^{(3)}(p' - p - q/2) G_E^{(s)}(Q^2)$,
with $G_E^{(s)}$ the nucleon's isoscalar electric form factor. Apart from tiny effects $\sim 1/M^2$ this is correct up to $O(eP^3)$ (counting $M \sim \Lambda_0$). 

The LO calculation of $G_C$ that uses this operator, together with the LO deuteron wave functions of Refs.~\cite{NTvK05,PVRA05}, provides a good description of the data out to $|{\bf q}| \approx 600$ MeV~\cite{PV08}. However, two-pion-exchange corrections affect the position of the form-factor minimum, and so must be considered for a precision description. If the potential at $O(P^3)$ is employed to calculate wave functions then the charge operator should be evaluated up to $O(e P^3)$. 
At that order $J_0$ contains a two-body contribution of one-pion range. The coefficient of this contribution is, however, determined by $g_A$, $f_\pi$, and the nucleon mass~\cite{Ph03}, and so results for $G_C$ and $G_Q$ are still predictions. 

\begin{figure}[th]
\includegraphics[width=2.4in]{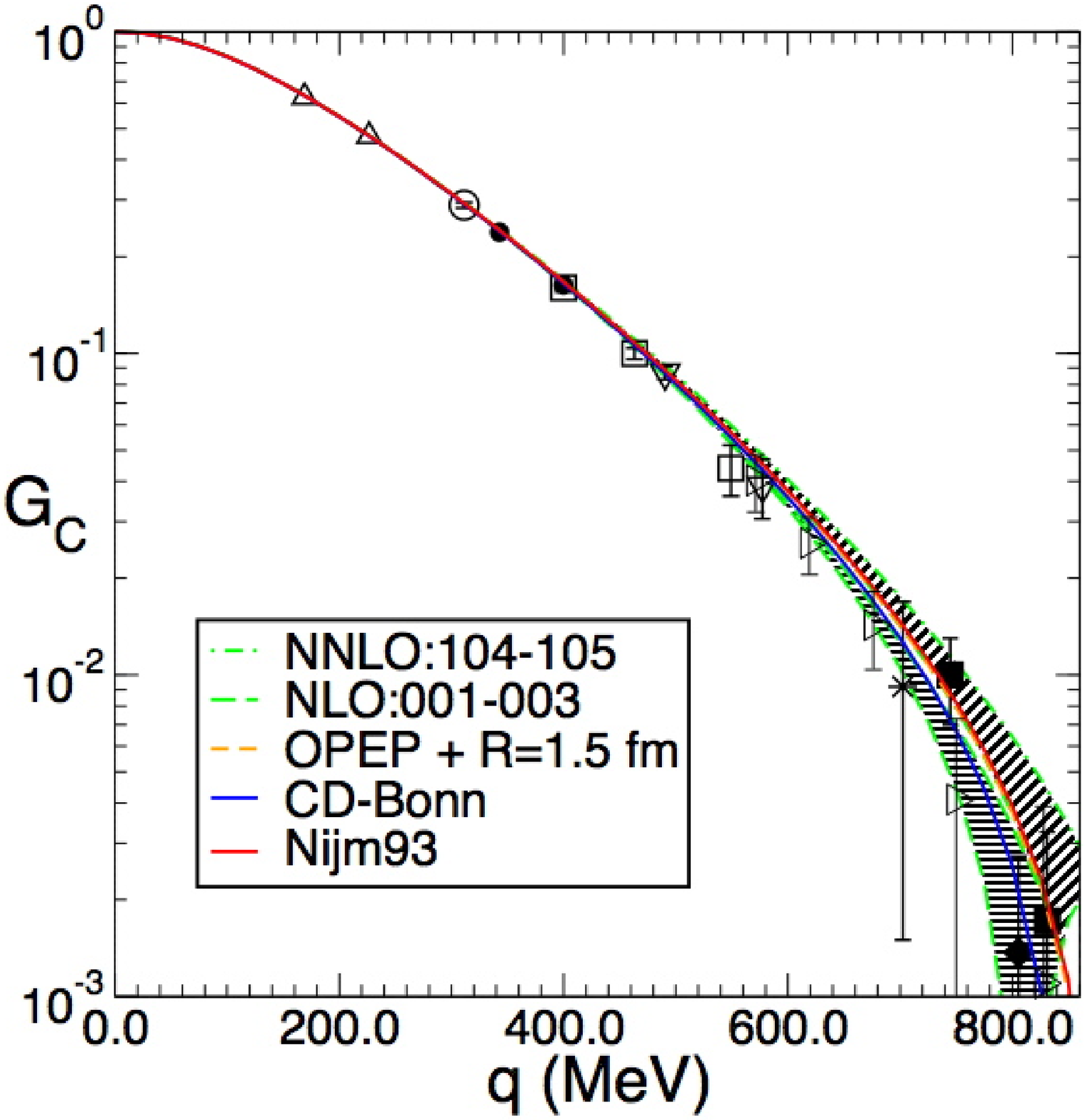}
\includegraphics[width=2.4in]{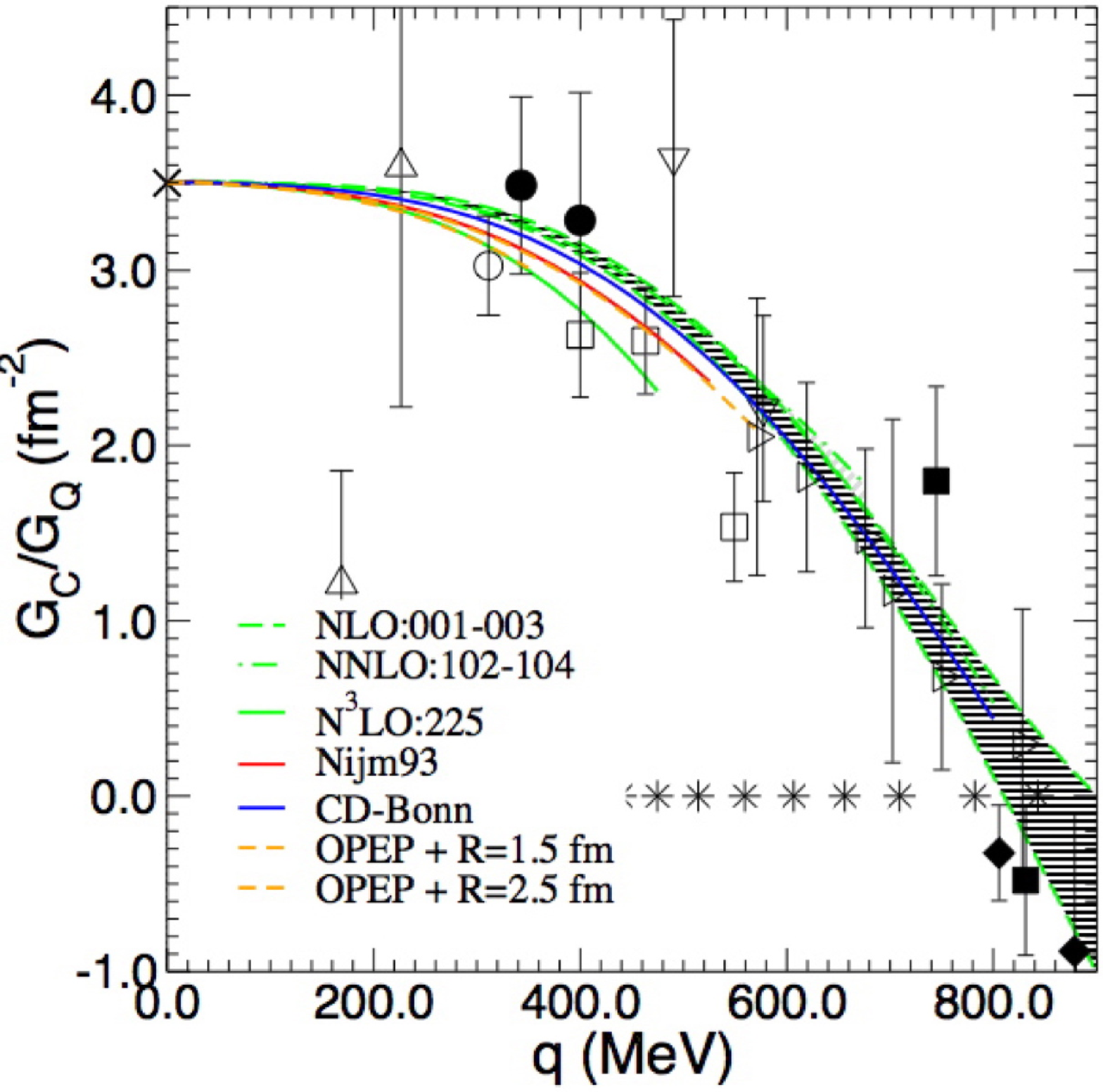}
\caption{Predictions for $G_C$ and $G_C/G_Q$ with various (low-cutoff) wave functions, most of which include two-pion exchange. The bands represent the theoretical uncertainty and the stars are the values of $|{\bf q}|$ where BLAST has data. The data shown is from Ref.~\cite{data}. Figures from Ref.~\cite{Ph07}, which should be consulted for details.}
\label{fig-GCGQ}
\end{figure}

The left panel of Fig.~\ref{fig-GCGQ} shows
that prediction for $G_C$ for various wave functions. The consistent, $O(eP^3)$, $\chi$ET result is obtained with 
the "NNLO" 
wave function, and is indicated by the diagonally shaded band. The variability comes from changes in the result as $\Lambda$ is varied. In these calculations we keep $\Lambda < 1$ GeV, which ensures that the $O(P^2)$ and $O(P^3)$ corrections to the deuteron wave function are only small perturbations to the LO result. The $O(eP^3)$ $\chi$ET result is, though, in markedly better agreement with the electron-deuteron scattering data than the LO $[O(e)]$ one.

In the $M \rightarrow \infty$ limit there are no two-pion-range two-body contributions to the isoscalar part of $J_0$ at $O(e P^4)$~\cite{EK09,Pa10} or $O(eP^5)$~\cite{Pa10}. There are also no one-pion-range pieces of $J_0^{(s)}$ at $O(eP^4)$~\cite{Pa10}. At $O(eP^5)$ a short-distance operator $\sim (N^\dagger N)^2 \nabla^2 A_0$ can affect the predictions for $G_C$. The coefficient of this operator is constrained by measurements of the deuteron charge radius. The NNLO wave functions of Ref.~\cite{Ep05} with the $O(e P^4)$ $J_0$ yield $\langle r^2 \rangle^{1/2}_{\rm pt}=1.975(1)$ fm. The uncertainty comes from short-distance differences between the wave functions. The deuteron isotope shift gives a value $1.9748(7)$ fm for the same quantity~\cite{Friar}. The very small difference between these two numbers constrains the LEC multiplying the $O(eP^5)$ short-distance contribution to $J_0^{(s)}$. This indicates that the prediction for $G_C$ shown in Fig.~\ref{fig-GCGQ} is accurate to $\pm 1.5$\% at $Q^2=0.16$ GeV$^2$~\cite{Pa10}. This claim will be tested by forthcoming JLab data.

The situation is analogous, but plays out slightly differently, in the case of $G_C/G_Q$. At $O(eP^3)$ the quadrupole moment, $Q_d$,
is quite sensitive to the short-distance physics included in the deuteron wave function. It varies
by 2\% when $\Lambda$ is changed from 500 to 800 MeV. Intriguingly, this is roughly the size of the
discrepancy between the $Q_d$ predicted at $O(eP^3)$ and the
experimental value $Q_d=0.2859(3)~{\rm fm}^2$. Ref.~\cite{Ph07} considered the short-distance operator at $O(eP^5)$ that represents the
contribution of modes above $\Lambda_0$ to $G_Q$. 
The LEC multiplying this short-distance $O(eP^5)$ operator
was fixed by demanding that its coefficient be such that the experimental $Q_d$
is reproduced. Thus, at $O(eP^5)$ $\chi$ET cannot predict $G_Q$ at $Q^2=0$,
but it {\it can} predict the $Q^2$-dependence of $G_Q$. That
prediction's theoretical uncertainty comes from the
$Q^2$-dependence of short-distance NN physics, and is $\sim 3$\% at $Q^2=0.16~{\rm GeV}^{2}$. The inclusion of this
short-distance, quadrupole operator leads to 
$\chi$ET predictions for $G_C/G_Q$ that have a different
$Q^2$-dependence to that obtained in potential models. BLAST data will provide a
significant test of this result.



\begin{theacknowledgments}
This research was supported by the U.~S. Department
of Energy, Office of Nuclear Physics, under contract No. DE-FG02-93ER40756 with Ohio University. I thank Charlotte Elster, Norbert Kaiser, Tae-Sun Park, Jerry Yang, and Evgeny Epelbaum, Jim Friar, and Stefan K\"olling for fruitful collaboration and discussions on the topics presented here. I am also grateful to the organizers of MENU for putting together a broad and stimulating program. 
\end{theacknowledgments}




\bibliographystyle{aipproc}   

\bibliography{sample}

\IfFileExists{\jobname.bbl}{}
 {\typeout{}
  \typeout{******************************************}
  \typeout{** Please run "bibtex \jobname" to optain}
  \typeout{** the bibliography and then re-run LaTeX}
  \typeout{** twice to fix the references!}
  \typeout{******************************************}
  \typeout{}
 }

\end{document}